\documentclass[12pt,preprint]{aastex}

\shorttitle{Hydrocarbon anions in interstellar clouds}

\shortauthors{Millar et al.}

\begin{document}




\title{Hydrocarbon anions in interstellar clouds and circumstellar envelopes}







\author{T. J. Millar, C. Walsh and M. A. Cordiner}

\affil{Astrophysics Research Centre, School of Mathematics and Physics, Queen's University Belfast, Belfast BT7 1NN, Northern Ireland}
\email{Tom.Millar@qub.ac.uk, cwalsh13@qub.ac.uk, m.cordiner@qub.ac.uk}

\and

\author{R. N\'{i} Chuim\'{i}n}

\affil{Jodrell Bank Centre for Astrophysics, School of Physics and Astronomy, University of Manchester, Sackville St., Manchester M60 1QD, UK}
\email{Roisin.Ni-Chuimin@manchester.ac.uk}

\and

\author{Eric Herbst}

\affil{Departments of Physics, Chemistry and Astronomy, The Ohio State University,
  Columbus, OH 43210-1106}
\email{herbst@mps.ohio-state.edu}


















\begin{abstract}
The recent detection of the hydrocarbon anion C$_6$H$^-$ in the interstellar 
medium has led us to investigate the synthesis of hydrocarbon anions in a 
variety of interstellar and circumstellar environments. We find that the anion/neutral 
abundance ratio can be quite large, on the order of at least a few percent, once the neutral 
has more than five carbon atoms. Detailed modeling shows that the column 
densities of C$_6$H$^-$ observed in IRC+10216 and TMC-1 can be reproduced. 
 Our calculations also predict that other hydrocarbon anions, such as 
C$_{4}$H$^{-}$ and C$_8$H$^-$, are viable candidates for detection in 
IRC+10216, TMC-1 and photon-dominated regions such as the Horsehead Nebula. 
\end{abstract}

























\section{Introduction}
The recent detection of C$_6$H$^-$ in the carbon-rich AGB star 
\object{IRC+10216} and the cold, dense interstellar cloud \object{TMC-1} by
\citet{mcc06} with abundance ratios relative to C$_6$H of
0.01--0.1 indicates that anions may play a more significant role in
interstellar physics and chemistry than heretofore believed.

The possibility that a relatively large fraction of molecular material in 
interstellar clouds might be in the form of anions was first suggested by 
\citet{her81} who pointed out that carbon chain molecules and other radicals have large 
electron affinities, leading to high radiative attachment rates such as those  measured by \cite{wood80} with the attendant possibility of
anion/neutral fractions on the order of
a few percent.  More recently, it has been recognized that PAH anions could
soak up a significant fraction of the free electrons in interstellar clouds 
and alter the charge balance to a significant degree \citep{lep88a} as well as 
providing significant heating through photodetachment of electrons in 
diffuse clouds \citep{lep88b}. \citet{pet96} has investigated the 
synthesis of CN$^-$ by the dissociative attachment of MgCN and MgNC and other
mechanisms,
while \citet{pet97} showed that C$_3$N$^-$ could be
detectable in interstellar clouds.  The study of large hydrocarbons anions 
received 
some attention following the observation by \citet{tul98} that 
absorption bands in several carbon-chain anions coincided with several of the 
diffuse interstellar bands. Subsequently, the formation of such species in 
diffuse clouds was studied by \citet{ruf99} who showed, in 
particular, that C$_7^-$ was unlikely to be a source for any DIBs. 
\citet{mil00} considered the formation of hydrocarbon anions 
containing more than six carbon atoms in the carbon-rich circumstellar 
envelope of \object{IRC+10216} and showed that appreciable column densities 
could arise in the outer envelope.

\section{Chemical Model}

\label{chem}

The basic route to the formation of anions is electron radiative attachment:
\begin{equation}  X + e^- \longrightarrow X^- + h\nu,  \end{equation} which has
been discussed in the context of bare carbon chains by \citet{ter00}.  The rate
coefficients of relevant processes have not been experimentally determined at low
temperatures and in this work we have used phase-state theory assuming $s$-wave
attachment with radiative stabilisation occurring by vibrational and electronic
transitions \citep{ter00}. Table \ref{tab:rates} contains some relevant 
attachment rate coefficients. The 
calculations of rates for these and other processes will be discussed in a 
separate paper (Herbst, in preparation). For C$_{n}$H radicals, we find that radiative attachment occurs at the
collisional rate once the number of carbon atoms is larger than five; for
C$_4$H, the attachment efficiency is only around 1\%.  \citet{bar01} have
shown experimentally that neither carbon-chain anions nor hydrocarbon anions
react with H$_2$ but do so with H atoms. We have also included loss reactions
of anions with C, C$^+$, C$_2$H$_2^+$ (in IRC+10216), and photons, with all cation-anion rate coefficients taken to be 10$^{-7}$ cm$^3$ s$^{-1}$.  
For the photodetachment rates, some of which are shown in Table \ref{tab:rates}, we have assumed the cross section $\sigma$ to depend on photon energy $\epsilon$ via the relation
\begin{equation}
\sigma =  \sigma_{0} (1 - EA/\epsilon)^{0.5},  \epsilon \geq EA,
\end{equation}
where $\sigma_{0} = 1.0 \times 10^{-17}$ cm$^{2}$ and 
 $EA$ is the electron affinity of
the  neutral molecule. Relevant electron affinities have been obtained from theoretical and experimental results (Herbst, in preparation). The
UV radiation field as described by \citet{mat83} has been adopted. We have extended earlier
models such as that by \citet{ruf99}, which included C$_{n}^{-}$, $n$ = 7--23,
and by \citet{mil00}, which included  C$_n$H$^-$, $n$ = 7--23, down to anions
containing four carbon atoms following the determination of the microwave
spectrum of C$_4$H$^-$ by \citet{gup07}.

In addition to the reactions discussed above, we have included
in the CSE and PDR models radiative association between carbon-chain anions and neutrals leading to carbon chains of up to 23 carbon atoms \citep{mil00}, 
 while in the dark cloud models we have included reactions 
between anions and O atoms, with a rate coefficient of 10$^{-10}$ cm$^3$ 
s$^{-1}$, unless specifically measured (Eichelberger et al. 2002).

\section{Results}
\label{results}

\subsection{Circumstellar Envelope}
\label{cses}

\begin{figure}
\includegraphics[angle=270, width=\textwidth]{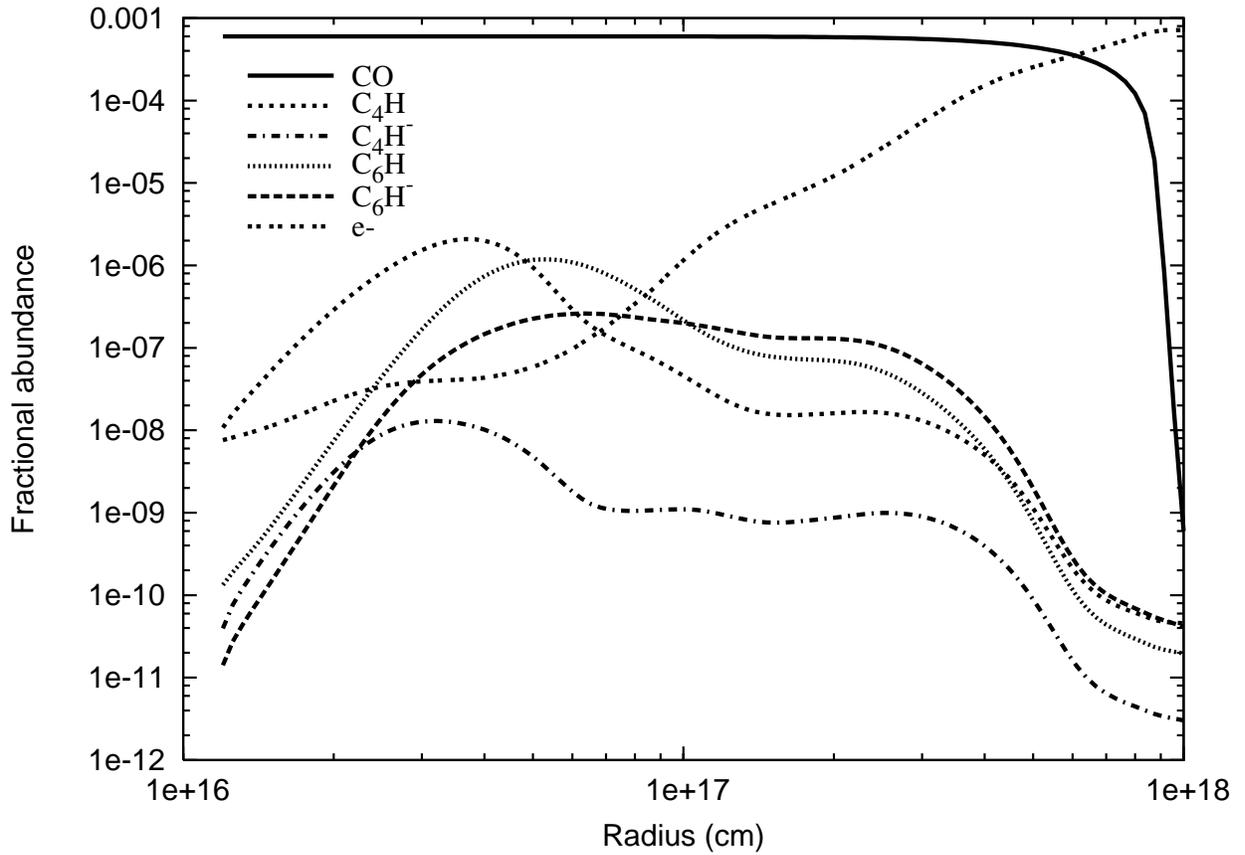}
\caption{IRC+10216 circumstellar chemistry model results of hydrocarbon and anion abundances (relative to the total number density), calculated as a function of distance from the center of the star.\label{fig:cse}}
\end{figure}

We model the carbon-rich AGB star IRC+10216 using the numerical 
model developed by \citet{mil00} and \citet{mil03}. They modelled the 
formation of anions with more than six carbon atoms and showed that the anion 
radial distribution could extend beyond 10$^{17}$ cm and relative to their neutral
analogues, abundance ratios approaching 0.1 (for C$_7^-$, for  example) could
be achieved. Some results of our present calculation in which we have extended our 
species down to C$_4$H$^-$ are shown in Table 
\ref{tab:cse} and Figure \ref{fig:cse}.  Our results show that column 
densities of $1.0\times10^{13}$ cm$^{-2}$ for C$_4$H$^-$ to $2.3\times10^{13}$ cm$^{-2}$
for C$_{10}$H$^{-}$ are achievable, with anion-to-neutral column density ratios 
of 0.008 for C$_4$H to around 0.3--0.4 for C$_6$H, C$_8$H and C$_{10}$H.  Note that although 
the electron attachment efficiency for C$_4$H is only around 1\% that of 
C$_6$H and C$_8$H, C$_4$H$^-$ has a relatively large column density since 
C$_4$H is more abundant.  Both C$_4$H$^-$ and C$_8$H$^-$ are candidates for 
detection in IRC+10216. In the region interior to a radial distance of $8\times10^{16}$ cm, the main loss of
anions occurs through reaction with H atoms, while mutual neutralisation with 
C$^+$ dominates in $(1.8-13)\times10^{17}$ cm. Elsewhere photodetachment is
the major loss route.

\citet{mcc06} measured a C$_6$H$^-$ column density of $3\times10^{12}$
cm$^{-2}$, approximately 60 times less than is produced in our model. Although 
the column density is over-produced, it is important to note that the
column densities of the large hydrocarbon chains are very sensitive to the
initial abundance of acetylene adopted in the model. This results from the
fact
that the most efficient method of growth is via the addition of C$_2$
units. For example, if the initial abundance of C$_2$H$_2$ is reduced by five,
the column density of C$_6$H is reduced by about 45 and that of C$_{10}$H by 100. The anion column densities
are not affected as much, typically reducing by about an order of magnitude, since less acetylene also leads to a reduction in the abundances of H atoms and C$^+$ in the envelope.  The observed anion-to-neutral ratio is 0.01--0.1 where the 
range is due to the range of reported C$_6$H column densities
\citep{kaw95,gue97}, but is in reasonable agreement with the calculated value
of 0.3, given the uncertainties in the rates of anion formation (by radiative
electron attachment), and destruction.

Finally we note that the anion abundances can be greater than that of
  free electrons in the region around $5\times10^{16}$ cm. This reflects the
  efficiency of carbon-chain growth and electron attachment. A similar result
  is found in dark cloud models when PAHs are included \citep{lep88a}.

\subsection{Dark Clouds}
\label{clouds}

\begin{figure}
\includegraphics[angle=270, width=\textwidth]{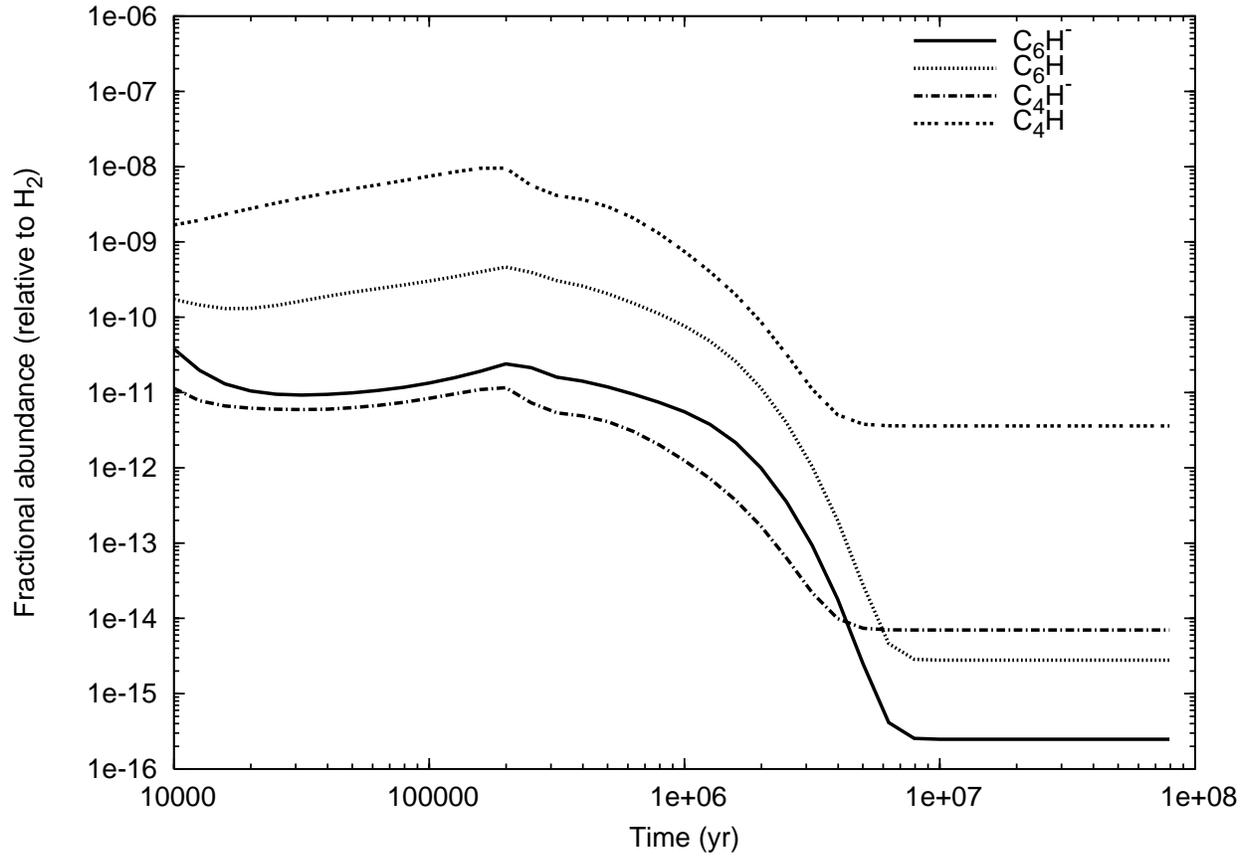}
\caption{TMC-1 dark cloud chemistry model results of hydrocarbon and anion abundances 
(relative to H$_2$ density), calculated as a function of time.\label{fig:tmc1}}
\end{figure}

Here we adopt parameters applicable to the cold dust cloud TMC-1 ($n({\rm H_2})
= 2\times10^{4}$ cm$^{-3}$, T = 10 K, A$_{V}$ = 10 mag.) and perform 
quasi-time-dependent modeling. 
The initial elemental abundances of C, N and O relative to H are 
$7.30\times10^{-5}$, $2.14\times10^{-5}$ and $1.76\times10^{-4}$, 
respectively.
Figure~\ref{fig:tmc1} shows the time-dependent evolution of the fractional
abundances (with respect to H$_{2}$) of C$_{4}$H$^{-}$ and C$_{6}$H$^{-}$ and
their corresponding neutrals from $10^{4}$ yr while  Table~\ref{tab:tmc1}
and Table~\ref{tab:tmc2} present anion column densities and anion/neutral 
ratios respectively, at both early-time ($3.16\times10^{5}$ yr) and steady 
state ($>10^8$ yr). We adopt N(H$_2$) = 10$^{22}$ cm$^{-2}$ in calculating the 
molecular column densities.

Steady-state abundances are very low due to the incorporation of carbon into 
CO but much better agreement occurs at earlier times in the evolution. The 
column density of C$_6$H$^-$ produced in our model at early time
($3.16\times10^{5}$ yr),  $1.35\times10^{11}$ cm$^{-2}$, is very close to
that observed by \citet{mcc06}, $1\times10^{11}$ cm$^{-2}$.   We calculate the
column density of the neutral C$_6$H at early time to be $2.58\times10^{12}$ 
cm$^{-2}$,
which  is in good agreement with the observed value of $4.10\times10^{12}$
cm$^{-2}$ \citep{bel99}.  The observed anion-to-neutral ratio is 0.025, which
is in very good agreement with the calculated early-time value of 0.052,
given the uncertainties in assorted rate coefficients.

\subsection{Photon-Dominated Regions}
\label{pdrs}

\begin{figure}
\includegraphics[angle=270, width=\textwidth]{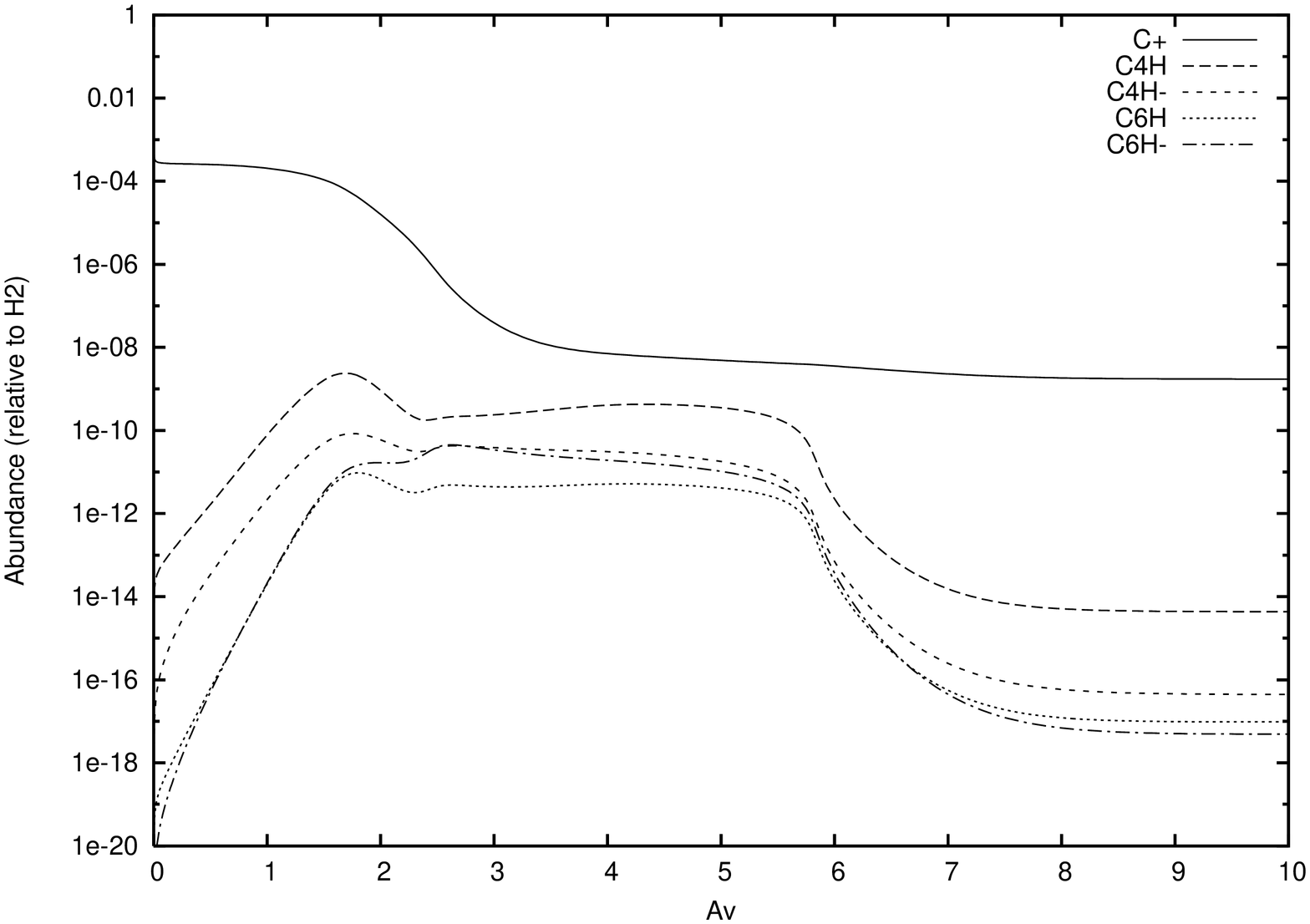}
\caption{Horsehead Nebula PDR chemistry model results of hydrocarbon and anion abundances (relative to H$_2$), calculated as a function of A$_V$.\label{fig:pdr}}
\end{figure}

We have used the code developed by the Meudon group \citep{pet06} with the
additional hydrocarbon chemistry described above. For this model we assume a
fixed temperature of 50~K and parameters representative of those in the
Horsehead Nebula; i.e.,  G = 60G$_0$, where G$_0$ is the interstellar UV radiation
field, A$_V$ = 10 mag. to the cloud center, and a total hydrogen density of
$2\times10^{4}$ cm$^{-3}$. These models are steady-state, which is achieved
rapidly due to the enhanced UV field and fast photodestruction rates. The PDR
models show that the anionic species formed can survive exposure to a high
radiation field present in regions such as the Horsehead Nebula because, given 
the high electron abundance,
 the formation path of electron attachment is efficient enough
to overcome the rapid photodetechment of the fragile anions.  This high 
electron abundance at low to intermediate extinction can be inferred from the 
C$^+$ abundance in Fig.~\ref{fig:pdr}, which also shows that significant
abundances of the anions arise at low extinction (A$_V$ $\sim$ 1.5--3 mag.). 
At low A$_V$, $\sim$ 0.5--2 mag., the dominant loss of anions is through
mutual neutralisation with C$^+$ (and to a lesser extent through reactions 
with atomic hydrogen). Thus, since n(C$^+$) = n(e) in this region, the anion/neutral abundance ratio depends on the
ratio of the rate coefficients for electron attachment and mutual
neutralisation. For species with more than five carbon atoms, this leads to
abundance ratios greater than unity; for smaller species, the ratio is less
than one.
In fact, the abundances of the 
anions are, in some cases, larger than the 
neutral species. Relevant results are listed in Table~\ref{tab:pdr}.
Most notably, C$_8$H$^-$ and C$_{10}$H$^-$ are more dominant than
the neutral species with fractional abundances over an order of 
magnitude greater than those of C$_8$H and C$_{10}$H.

Since the PDR model provides abundances perpendicular to the line-of-sight
for a PDR seen edge on, as is the case for the Horsehead Nebula, it is more appropriate to compare fractional abundances rather than column densities. The peak abundance of C$_4$H is consistent with observations of the
Horsehead region made by \citet{tes04}. Although the relative abundance of
C$_4$H$^-$ (3.5\%) to its neutral analog is not as high as C$_6$H$^-$ or C$_8$H$^-$, the higher
abundance of the neutral species allows a significantly large absolute value of
this anion to be formed.
The C$_6$H$^-$ anion has a peak abundance of 4.7 times that of the neutral 
species. The detection of C$_6$H in the Horsehead Nebula was reported by 
\citet{tes04}, who calculated a peak abundance relative to H$_2$ of 
$1.0\times10^{-10}$, about one order of magnitude greater than calculated. 
It is possible that we are missing an important formation mechanism in 
this environment.  One possibility is that hydrocarbon
abundances are boosted by destruction of PAH particles, rather than being
synthesized from smaller species. In any case, our anion/neutral ratios are
independent of the mode of formation of the neutral.

\section{Conclusion}
We find that electron attachment to hydrocarbon molecules is very efficient 
for species containing more than five carbon atoms. 
The anions are created so efficiently that they can be abundant even in regions
where they are destroyed rapidly by photons, such as CSEs and PDRs, and 
in dark clouds, despite the 
low fractional abundance of electrons. In particular, we find that anions such 
as C$_6$H$^-$, C$_8$H$^-$ and C$_{10}$H$^-$ can have abundances relative to their neutral 
analogs ranging from a few percent to greater than  unity,
while C$_4$H$^-$, although formed relatively inefficiently through
electron attachment, can be observable in astronomical objects, given the large
column densities of C$_4$H detected.  In fact, as this paper was being 
prepared, we became aware of a preprint by Cernicharo et al. in which the 
detection of C$_{4}$H$^{-}$ in IRC+10216 is reported with a 
C$_4$H$^-$/C$_6$H$^-$
column density ratio of 1/7,  compared to our calculated value of 1/17.

\acknowledgments
Astrophysics at QUB is supported by a grant from PPARC. RNC and TJM are 
supported by by the European Community's human potential programme under
contract MCRTN-CT-512302, `The Molecular Universe', CW by a scholarship from 
the Northern Ireland Department of Employment and Learning, and MAC by QUB. EH 
acknowledges support for his research program in astrochemistry by the National
Science Foundation.  We thank Dr. Aigen Li for providing us his code to 
calculate radiation intensity as a function of wavelength and extinction.























\clearpage
\begin{deluxetable}{lllllll}
\tabletypesize{\scriptsize}
\tablecaption{Some reactions and rate coefficients\tablenotemark{a,b}\label{tab:rates}}
\tablewidth{0pt}
\tablehead{
\colhead{R$_1$} & \colhead{R$_2$} & \colhead{P$_1$} & \colhead{P$_2$} & \colhead{$\alpha$} &
\colhead{$\beta$} & \colhead{$\gamma$} 
}
\startdata
e- & C4H & C4H- &  & 2.00(-9) & -0.5 &  \\
e- & C5  & C5-  &  & 3.30(-8) & -0.5 &  \\
e- & C5H & C5H- &  & 9.00(-10)& -0.5 &  \\
e- & C6  & C6-  &  & 1.70(-7) & -0.5 &  \\
e- & C6H & C6H- &  & 6.00(-8) & -0.5 &  \\
e- & C7  & C7-  &  & 5.00(-7) & -0.5 &  \\
e- & C7H & C7H- &  & 2.00(-7) & -0.5 &  \\
e- & C8  & C8-  &  & 1.70(-7) & -0.5 &  \\
e- & C8H & C8H- &  & 6.00(-8) & -0.5 &  \\
e- & C9  & C9-  &  & 5.00(-7) & -0.5 &  \\
e- & C9H  & C9H-  &  & 2.00(-7) & -0.5 &  \\
e- & C10  & C10-  &  & 1.70(-7) & -0.5 &  \\
e- & C10H & C10H- &  & 6.00(-8) & -0.5 &  \\
C4H- & PHOTON & C4H  & e- & 1.80E-09 &  0.0 & 2.0 \\
C5-  & PHOTON & C5   & e- & 3.00E-09 &  0.0 & 1.5 \\
C5H- & PHOTON & C5H  & e- & 3.70E-09 & 0.0  & 1.5 \\
C6-  & PHOTON & C6   & e- & 1.30E-09 & 0.0  & 2.0 \\
C6H- & PHOTON & C6H  & e- & 1.50E-09 & 0.0  & 2.0 \\
C7-  & PHOTON & C7   & e- & 2.40E-09 & 0.0  & 1.5 \\
C7H- & PHOTON & C7H  & e- & 3.00E-09 & 0.0  & 1.5 \\
C8-  & PHOTON & C8   & e- & 1.10E-09 & 0.0  & 2.0 \\
C8H- & PHOTON & C8H  & e- & 1.40E-09 & 0.0  & 2.0 \\
C9-  & PHOTON & C9   & e- & 1.60E-09 & 0.0  & 2.0 \\
C9H- & PHOTON & C9H  & e- & 2.00E-09 & 0.0  & 2.0 \\
C10-  & PHOTON & C10   & e- & 1.00E-09 & 0.0  & 2.0 \\
C10H- & PHOTON & C10H  & e- & 1.40E-09 & 0.0  & 2.0 \\
\enddata
\label{rates}
\tablenotetext{a} {Rate coefficients $ k = \alpha (T/300)^{\beta}\exp{(-\gamma/T)}$ in units of
cm$^{3}$ s$^{-1}$}
\tablenotetext{b}{Herbst, in preparation.}
\end{deluxetable}

\clearpage

\begin{deluxetable}{lccc}
\tablecaption{IRC+10216 hydrocarbon anion chemistry model results\label{tab:cse}}
\tablewidth{0pt}
\tablecolumns{4}
\tablehead{
\colhead{Species} & \colhead{Column density} & \colhead{Peak  abundance} &
\colhead{Observed column density}\\
 & \colhead{(cm$^{-2}$)} & rel. to H$_{2}$ &
\colhead{(cm$^{-2}$)}
}
\startdata
C$_4$H     & $1.3\times10^{15}$ &  $2.1\times10^{-6}$ & $2-9\times10^{15}$ (1, 2, 3, 4)\\
C$_4$H$^-$ & $1.0\times10^{13}$ &  $1.3\times10^{-8}$ & \\ 
C$_6$H     & $5.7\times10^{14}$ &  $1.2\times10^{-6}$ & $0.3-3\times10^{14}$ (3, 4)\\
C$_6$H$^-$ & $1.7\times10^{14}$ &  $2.6\times10^{-7}$ & $3\times10^{12}$ (5)\\
C$_8$H     & $2.1\times10^{14}$ &  $4.4\times10^{-7}$ & $5\times10^{12}$ (4)\\
C$_8$H$^-$ & $5.8\times10^{13}$ &  $1.4\times10^{-7}$ & \\
C$_{10}$H  & $5.8\times10^{13}$ &  $1.3\times10^{-7}$ & \\
C$_{10}$H$^-$ & $2.3\times10^{13}$ & $8.3\times10^{-8}$ & \\
\enddata
\tablecomments{References.-- (1) \citet{ave92}, (2) \citet{day93}, (3) \citet{kaw95}, (4) \citet{gue97}, (5) \citet{mcc06}.}
\end{deluxetable}

\clearpage

\begin{deluxetable}{lccc}
\tablecaption{TMC-1 hydrocarbon anion chemistry model results \label{tab:tmc1}}
\tablewidth{0pt}
\tablecolumns{4}
\tablehead{
 & \colhead{Early time} & \colhead{Steady state} \\
\colhead{Species} & \colhead{Column density} & \colhead{Column density} & \colhead{Observed column density}\\
 & \colhead{(cm$^{-2}$)} & \colhead{(cm$^{-2}$)} & \colhead{(cm$^{-2}$)}\\
 }
\startdata
C$_4$H     & $4.30\times10^{13}$ & $3.60\times10^{10}$ & $3.4\times10^{14}$ (1)\\
C$_4$H$^-$ & $5.58\times10^{10}$ & $6.99\times10^{7}$ & \\ 
C$_6$H     & $2.58\times10^{12}$ & $2.78\times10^{7}$ & $4.1\times10^{12}$ (2)\\
C$_6$H$^-$ & $1.35\times10^{11}$ & $2.47\times10^{6}$ & $1.0\times10^{11}$ (3)\\
C$_8$H     & $7.76\times10^{11}$ & $1.63\times10^{6}$ & $2.2\times10^{11}$ (2)\\
C$_8$H$^-$ & $3.27\times10^{10}$ & $6.83\times10^{4}$ & \\
C$_{10}$H  & $8.98\times10^{10}$ & $3.78\times10^{3}$ & \\
C$_{10}$H$^-$ & $3.67\times10^{9}$ & $2.05\times10^{2}$ & \\
\enddata
\tablecomments{References: (1) \citet{gue82}, (2) \citet{bel99}, (3) \citet{mcc06}.}
\end{deluxetable}

\clearpage

\begin{deluxetable}{lccc}
\tablecaption{TMC-1 hydrocarbon anion-to-neutral ratios C$_{n}$H$^{-}$/C$_{n}$H \label{tab:tmc2}}
\tablewidth{0pt}
\tablecolumns{4}
\tablehead{
\colhead{n} & \colhead{Early time} & \colhead{Steady state} & \colhead{Observed}
 }
\startdata
4 & $1.30\times10^{-3}$ & $1.94\times10^{-3}$ & \\
6 & $5.23\times10^{-2}$ & $8.91\times10^{-2}$ & $2.50\times10^{-2}$ (1)\\ 
8 & $4.21\times10^{-2}$ & $5.42\times10^{-2}$ & \\
10& $4.10\times10^{-2}$ & $5.42\times10^{-2}$ & 
\enddata
\tablecomments{References: (1) \citet{mcc06}.}
\end{deluxetable}

\clearpage

\begin{deluxetable}{lcccc}
\tablecaption{Horsehead Nebula hydrocarbon anion chemistry model results\label{tab:pdr}}
\tablewidth{0pt}
\tablecolumns{3}
\tablehead{
\colhead{Species} & \colhead{Peak abundance} & \colhead{Obs. peak abundance} \\
& rel. to H$_{2}$ & rel. to H$_{2}$ 
}
\startdata
C$_4$H     & $2.4\times10^{-9}$ &  $3.0\pm 1.2\times10^{-9}$ (1)\\
C$_4$H$^-$ & $8.4\times10^{-11}$ &  \\
C$_6$H     & $9.6\times10^{-12}$ &  $1.0\pm 0.6\times10^{-10}$ (1)\\
C$_6$H$^-$ & $4.5\times10^{-11}$ &  \\
C$_8$H     & $5.3\times10^{-12}$ &  \\
C$_8$H$^-$ & $9.3\times10^{-11}$ &  \\
C$_{10}$H  & $3.8\times10^{-12}$ &  \\
C$_{10}$H$^-$ & $5.5\times10^{-11}$ & \\
\enddata
\tablecomments{References: (1) \citet{tes04}.}
\end{deluxetable}

\clearpage

\clearpage

\clearpage

\begin{thebibliography}{}
\bibitem[Avery et al. (1992)]{ave92} Avery, L. W., Amano, T., Bell, M. B., 
  Feldman, P. A., Johns, J. W. C., MacLeod, J. M., Matthews, H. E., Morton, D. 
  C., Watson, J. K. G., Turner, B. E., Hayashi, S. S., Watt, G. D., \& 
  Webster, A. S. 1992 \apjs, 83, 363 
\bibitem[Barckholtz et al.(2001)]{bar01} Barckholtz, C., Snow, T. P., \& 
  Bierbaum, V. M. 2001, \apjl, 547, L171
\bibitem[Bell et al. (1999)]{bel99} Bell, M. B., Feldman, P. A., Watson, J. K. 
  G., McCarthy, M. C., Travers, M. J., Gottlieb, C. A., \& Thaddeus, P. 1999, 
  \apj, 518, 740
\bibitem[Dayal \& Bieging (1993)]{day93} Dayal, A., \& Bieging, J. H. 1993, \apj,
  407, L37
\bibitem[Eichelberger et al. (2002)]{eic02} Eichelberger, B., Barckholtz, C., 
  Stepanovic, M., Bierbaum, V. M., \& Snow, T. P. 2002, NASA Laboratory 
  Astrophysics Workshop, ed. F. Salama, NASA/CP-2002-21186, p. 120 
\bibitem[Gu\'{e}lin et al. (1982)]{gue82} Gu\'elin, M., Friberg, P., \& 
  Mezaoui, A. 1982, \aap, 109, 23
\bibitem[Gu\'{e}lin et al. (1997)]{gue97} Gu\'elin, M., Cernicharo, J., 
  Travers, M. J., McCarthy, M. C., Gottlieb, C. A., Thaddeus, P., Ohishi, M., 
  Saito, S., \& Yamamoto, S. 1997, \aap, 317, L1 
\bibitem[Gupta et al.(2007)]{gup07} Gupta, H. C., Br\"unken, S., Tamassia, F.,
  Gottlieb, C. A., McCarthy, M. C., \& Thaddeus, P. 2007, \apjl, 655, L57
\bibitem[Herbst (1981)]{her81} Herbst, E. 1981, Nature, 289, 656
\bibitem[Kawaguchi et al (1995)]{kaw95} Kawaguchi K., Kasai, Y., Ishikawa, S., 
  \& Kaifu, N. 1995, \pasp, 47, 853
\bibitem[Le Petit et al.(2006)]{pet06} Le Petit, F., Nehme\'e, C., Le Bourlot,
  J., \& Roueff, E. 2006, \apjs, 164, 506
\bibitem[Lepp \& Dalgarno(1988a)]{lep88a} Lepp, S., \& Dalgarno, A. 1988a, 
  \apj, 324, 553
\bibitem[Lepp \& Dalgarno (1988b)]{lep88b} Lepp, S., \& Dalgarno, A. 1988b, 
  \apj, 335, 769
\bibitem[Mathis et al.(1983)]{mat83} Mathis, J. S., Mezger, P. G., \& 
   Panagia, N. 1983, \aap, 128, 212
\bibitem[McCarthy et al.(2006)]{mcc06} McCarthy, M. C., Gottlieb, C. A., 
  Gupta, H. C., \& Thaddeus, P. 2006, \apjl, 652, L141
\bibitem[Millar (2003)]{mil03} Millar, T. J. 2003, in Asymptotic Giant Branch 
  Stars, eds. H. J. Habing \& H. Olofsson, New York: Springer, 247
\bibitem[Millar et al.(2000)]{mil00} Millar, T. J., Herbst, E., \& Bettens, 
  R. P. A. 2000, \mnras, 316, 195
\bibitem[Petrie(1996)]{pet96} Petrie, S. 1996, \mnras, 281, 137
\bibitem[Petrie \& Herbst(1997)]{pet97} Petrie, S., \& Herbst, E. 1997, \apj,
  491, 210 
\bibitem[Ruffle et al.(1999)]{ruf99} Ruffle, D. P., Bettens, R. P. A., 
  Terzieva, R., \& Herbst, E. 1999, \apj, 523, 678 
\bibitem[Terzieva \& Herbst(2000)]{ter00} Terzieva, R., \& Herbst, E. 2000, 
  Int. J. Mass Spectrom., 201, 135
\bibitem[Teyssier et al.(2004)]{tes04} Teyssier, D., Foss{\'e}, D., Guerin, 
  M., Pety, J. Abergel, A., \& Roueff, E. 2004, \aap, 417, 135
\bibitem[Tulej et al.(1998)]{tul98} Tulej, M., Kirkwood, D. A., Pachkov, M., 
  \& Maier, J. P. 1998, \apjl, 506, L69
  \bibitem[Woodin et al.(1980)]{wood80} Woodin, R., Foster, M. S., \& Beauchamp, J. L. 1980, J. Chem. Phys., 72, 4223
\end{thebibliography}
\end{document}